\documentclass{mn2e}
\usepackage{psfig}

\newcommand{\etal}{{\it et al.}}
\newcommand{\gggga}{\mathrel{\mathchoice {\vcenter{\offinterlineskip\halign{\hfil
$\displaystyle##$\hfil\cr>\cr\sim\cr}}}
{\vcenter{\offinterlineskip\halign{\hfil$\textstyle##$\hfil\cr
>\cr\sim\cr}}}
{\vcenter{\offinterlineskip\halign{\hfil$\scriptstyle##$\hfil\cr
>\cr\sim\cr}}}
{\vcenter{\offinterlineskip\halign{\hfil$\scriptscriptstyle##$\hfil\cr
>\cr\sim\cr}}}}} 
\newcommand{\lllla}{\mathrel{\mathchoice {\vcenter{\offinterlineskip\halign{\hfil
$\displaystyle##$\hfil\cr<\cr\sim\cr}}}
{\vcenter{\offinterlineskip\halign{\hfil$\textstyle##$\hfil\cr
<\cr\sim\cr}}}
{\vcenter{\offinterlineskip\halign{\hfil$\scriptstyle##$\hfil\cr
<\cr\sim\cr}}}
{\vcenter{\offinterlineskip\halign{\hfil$\scriptscriptstyle##$\hfil\cr
<\cr\sim\cr}}}}} 

\title[Mission: Impossible]{Mission: Impossible (Escape from the Lyman Limit)
       \thanks{Based on observations taken with the NASA/ESA Hubble Space
       Telescope, which is operated by AURA under NASA contract NAS5-26555}}

\author[A. Fern\'andez-Soto et al.]
	{A.~Fern\'andez-Soto,$^{1,2}$\thanks{Marie Curie Fellow}\thanks{Email:
        alberto.fernandez@uv.es} K.M.~Lanzetta,$^3$ 
	H.-W.~Chen,$^{4,5}$\thanks{Hubble Fellow} \\
        $^1$Osservatorio Astronomico di Brera, Via Bianchi 46, Merate
        (LC), I-23807, Italy\\
        $^2$Observatori Astron\`omic, Universitat de Val\`encia, Burjassot 
        (Val\`encia), E-46100, Spain\\
        $^3$Department of Physics and Astronomy, State University of New 
        York at Stony Brook, Stony Brook, NY 11794-3800, U.S.A.\\
        $^4$The Observatories of the Carnegie Institution of Washington,
        813 Santa Barbara Street, Pasadena, CA 91101, U.S.A. \\
	$^5$Center for Space Research, Massachusetts Institute of Technology, 
	Cambridge, MA 02139-4307, U.S.A.}
\begin{document}

\date{Accepted ---. Received ---; in original form ---}

\pagerange{\pageref{firstpage}--\pageref{lastpage}} \pubyear{2002}

\maketitle

\label{firstpage}

\begin{abstract}
We investigate the intrinsic opacity of high-redshift galaxies to outgoing
ionising photons using high-quality photometry of a sample of 27
spectroscopically-identified galaxies of redshift $1.9 < z< 3.5$ in the
Hubble Deep Field.  Our measurement is based on maximum-likelihood fitting of
model galaxy spectral energy distributions---including the effects of
intrinsic Lyman-limit absorption and random realizations of intervening
Lyman-series and Lyman-limit absorption---to photometry of galaxies from
space- and ground-based broad-band images.  Our method provides several
important advantages over the methods used by previous groups, including most
importantly that two-dimensional sky subtraction of faint-galaxy images is
more robust than one-dimensional sky subtraction of faint-galaxy spectra.  We
find at the $3\sigma$ {\it statistical} confidence level that {\it on
average} no more than 4\,\% of the ionising photons escape galaxies of
redshift $1.9 < z< 3.5$.  This result is consistent with observations of low-
and moderate-redshift galaxies but is in direct contradiction to a recent
result based on medium-resolution spectroscopy of high-redshift ($z \approx
3$) galaxies. Dividing our sample in subsamples according to luminosity,
intrinsic ultraviolet colour, and redshift, we find no evidence for selection
effects that could explain such discrepancy. Even when all systematic effects
are included, the data could not realistically accomodate any escape fraction
value larger than $\approx$ 15\%.
\end{abstract}


\section{Introduction}

  The diffuse ultraviolet background is one of the key ingredients in the
recipe leading to galaxy formation (Miralda-Escud\'e \& Ostriker 1990; Giroux
\& Shapiro 1996).  The intensity of the diffuse ultraviolet background has
been estimated at moderate and high redshifts using the proximity effect on
QSO absorbers (Liske \& Williger 2001; Scott \etal\ 2001; Fern\'andez-Soto
\etal\ 1995; Kulkarni \& Fall 1993), and results obtained so far suggest that
it is too large to be explained by QSOs alone (Madau \& Shull 1996, but see
also e.g. Giallongo \etal\ 1996).  Although systematic errors may make these
measurements uncertain (Loeb \& Eisenstein 1995, Pascarelle \etal\ 2001),
several suggestions have been put forward to explain the observed extra
intensity.  One of the most popular suggestions calls for the escape of a
large fraction of ionising photons from high-redshift galaxies. If correct,
this would imply that the interstellar media of high-redshift galaxies must
be at least partially transparent to ionising photons.

  But the best measurements available at low and moderate redshifts show that
galaxies are (at least nearly) opaque to ionising photons.  The large 
column densities of neutral hydrogen that surround most galaxies should be 
enough, depending on how they are distributed, to very effectively quench any
outgoing flux of ionising photons.  In fact, Leitherer \etal\ (1995) find, 
in a sample of four starburst galaxies at an average redshift $\langle z 
\rangle \approx 0.02$, that the limits to the escape fraction of Lyman limit
photons range from $f_{\rm esc}<0.0095$ to $f_{\rm esc}<0.15$ (with an 
observed flux ratio at 900 \AA\ and 1500 \AA\ of $F_{9/15}<0.10 \relbar 0.20$).
Hurwitz \etal\ (1997) analyse the same data and obtain slightly less stringent 
limits, ranging from $f_{\rm esc}<0.032$ to $f_{\rm esc}<0.57$.  Deharveng 
\etal\ (2001) find a more stringent limit $f_{\rm esc}<0.064$ for a starburst 
galaxy of redshift $z=0.0448$.  At higher redshifts ($z \approx 1$), Ferguson 
(2001) finds similar upper limits $f_{\rm esc}<0.20$ to the escape fraction.

  It is much less clear whether galaxies at still higher redshifts are as
opaque to ionising photons as galaxies at low and moderate redshifts.
Accurate measurements become progressively more difficult at higher redshifts,
because galaxies are much fainter and the presence of Lyman
$\alpha$ forest absorption introduces further uncertainties.  The only two
measurements available so far are from Steidel, Pettini \& Adelberger (2001)
and Giallongo \etal\ (2002).  Adopting a correction factor for intervening
Lyman $\alpha$ absorption determined from a composite QSO
spectrum, Steidel \etal\ reported $F_{9/15}=0.22 \pm 0.05$ using 29 galaxies
at $\langle z \rangle = 3.40$, while Giallongo \etal\ reported a $1\sigma$
upper limit $F_{9/15}\la 0.05$ using two galaxies at $z\sim 3$.  The former
measurement indicates that galaxies at $z\sim 3$ are much more transparant to
ionising photons, contributing an equal amount of ionising flux to the
ultraviolet background radiation as the QSOs.  The latter measurement indicates
otherwise.

  In this article, we present a new measurement of $f_{\rm esc}$ using 27
galaxies at redshifts $1.9 < z < 3.5$ in the Hubble Deep Field (HDF).  All of
these galaxies have secure spectroscopic redshifts (see Cohen \etal\
2000--C00--and Dawson \etal\ 2001--D01) and accurate broad-band photometry
from HST/WFPC2 observations in the F300W, F450W, F606W, and F814W bandpasses.
Our measurement is based on deep space-based images, which provides two
important advantages over the methods used by previous groups: First, our sky
subtraction is more accurate, because two-dimensional sky subtraction in
faint-galaxy images is more robust than one-dimensional sky subtraction in
faint-galaxy spectra.  Second, the wavelength interval below the rest-frame
Lyman limit over which we integrate is larger, because the space-based
broad-band images are sensitive to observed-frame wavelengths as short as
$\lambda \approx 2800$ \AA, allowing us to integrate over rest-frame
wavelength intervals as large as $\approx 100$ \AA\ in some cases.  Our
method also improves upon previous work by accounting for the effect of the
Lyman $\alpha$ forest absorption by performing a large numer of random
realizations of a parameterized distribution, to account for variations of
the absorbing clouds across different lines of sight.

  Our results show that the escape fraction estimated from the ensemble of 27
galaxies has a (statistical) $3\sigma$ upper limit of $f_{\rm esc} \la
0.04$. Given that our galaxy sample spans a range of properties in redshift,
intrinsic colours, and luminosities, we also explore possible systematic
effects by dividing the galaxies into different subsamples. We do not find
any evidence to support that bluer galaxies have higher escape fractions of
ionizing photons.  Instead, we find that more luminous galaxies appear to
have shallower ultraviolet spectral slopes and higher escape fractions (at
the 2$\sigma$ level of significance). We do also search for other possible
systematic effects that could be inherent to our method, and find that other
effects cannot in any realistic way account for a large increase in the
escape fraction.

\section{Data}

  We consider all galaxies in the HDF with known spectroscopic redshifts in
the range $1.9 < z < 3.5$.  The lower limit is chosen so that the rest-frame
Lyman limit begins to affect the photometry in the F300W bandpass.  The upper
limit is chosen so that we can still reasonably distinguish absorption due to
the Lyman $\alpha$ forest from the intrinsic Lyman limit absorption.  We
previously measured high-quality photometry of these galaxies using HST/WFPC2
through the F300W, F450W, F606W, and F814W filters (Fern\'andez-Soto,
Lanzetta \& Yahil 1999--FLY99 hereafter; Lanzetta \etal\ in preparation).
The sample includes 27 galaxies, as given in Table 1. The reader is referred
to FLY99, C00, D01, and Fern\'andez-Soto \etal\ 2001 for more details on
individual galaxies.

\begin{table}
  \centering 
   \caption{Properties of the galaxy sample used in the analysis. Celestial
   coordinates are J2000.}
   \begin{tabular}{c c c c c}
\hline
  RA-12$^h$ & Dec-62$^o$ & FLY99 &  $z$  &  AB(814) \\ \hline 
   36:49.83 &   14:15.0  & 1016  & 1.980 &    23.40 \\ 
   36:48.32 &   14:16.6  & 1044  & 2.005 &    23.40 \\
   37:00.09 &   12:25.2  & 0048  & 2.050 &    23.77 \\
   36:54.73 &   13:14.8  & 0670  & 2.232 &    24.29 \\
   36:55.07 &   13:47.1  & 0831  & 2.233 &    24.52 \\
   36:50.11 &   14:01.1  & 0960  & 2.237 &    24.58 \\
   36:54.62 &   13:41.3  & 0806  & 2.419 &    25.27 \\
   36:45.88 &   14:12.1  & 1054  & 2.427 &    25.21 \\
   36:43.27 &   12:38.9  & 0664  & 2.442 &    24.81 \\
   36:53.19 &   13:22.7  & 0742  & 2.489 &    24.83 \\
   36:44.64 &   12:27.4  & 0517  & 2.500 &    23.73 \\
   36:41.72 &   12:38.8  & 0702  & 2.591 &    24.60 \\
   36:45.35 &   11:52.7  & 0175  & 2.799 &    23.24 \\
   36:44.09 &   13:10.8  & 0815  & 2.929 &    24.03 \\
   36:47.77 &   12:55.7  & 0687  & 2.931 &    23.95 \\
   36:46.93 &   12:26.1  & 0444  & 2.969 &    25.24 \\
   36:48.29 &   11:45.9  & 0021  & 2.980 &    25.07 \\
   36:53.43 &   13:29.4  & 0762  & 2.991 &    24.60 \\
   36:45.36 &   13:47.0  & 0964  & 3.160 &    25.09 \\
   36:51.20 &   13:48.8  & 0897  & 3.162 &    25.22 \\
   36:53.60 &   14:10.2  & 0955  & 3.181 &    24.55 \\
   36:41.23 &   12:02.9  & 0390  & 3.220 &    24.03 \\
   36:49.81 &   12:48.8  & 0568  & 3.233 &    25.18 \\
   36:52.99 &   14:08.5  & 0957  & 3.367 &    26.85 \\
   36:52.75 &   13:39.1  & 0825  & 3.369 &    25.07 \\
   36:52.41 &   13:37.8  & 0824  & 3.430 &    24.79 \\
   36:39.57 &   12:30.5  & 0688  & 3.475 &    25.40 \\ \hline
   \end{tabular}
\end{table}

    We determine the absolute $AB$ magnitude of each galaxy using a fiducial
spectral energy distribution determined from a template fitting technique
that we will describe in the next section.  We adopt a cosmological model
with vacuum energy density $\Omega_\Lambda=0.65$, matter density
$\Omega_M=0.35$, and a Hubble constant $H_0 = 65$ km s$^{-1}$ Mpc$^{-1}$.
Absolute $AB$ magnitudes of the galaxies at rest-frame wavelengths $\lambda
=1500$ \AA\ range from $\approx -21.8$ to $\approx -25.4$. Apart from these
absolute magnitudes, all other results in this paper are independent of the
chosen cosmological model.

\section{Method}

  Our method is based on maximum-likelihood fitting of model galaxy spectral
energy distributions, including the effects of intrinsic Lyman-limit absorption
and random realizations of intervening Lyman-series and Lyman-limit absorption,
to space- and ground-based broad-band images.  In practice, our method proceeds
in several steps as follows:

  First, we determine a fiducial spectral energy distribution of each galaxy by
fitting model spectral energy distributions to the broad-band photometric
measurements that fall between rest-frame wavelengths $\lambda = 1250$ and 2800
\AA.  The wavelength range is chosen to avoid the spectral discontinuity
produced by the Lyman series (rest-frame wavelength $\lambda < 1215$ \AA) and
the Balmer break (rest-frame wavelengths $\lambda \gggga$ 3000 \AA). The
fiducial spectral energy distributions are based on the six
spectrophotometric templates (of E, Sbc, Scd, Irr, SB1, and SB2 galaxies) of
our previous photometric redshift measurements (Yahata \etal\ 2000; see also
Ben\'{\i}tez 2000) together with a power-law form $f_\nu 
\propto\lambda^\alpha$.  (The power-law form is required because a few of the
galaxies are found to be bluer--in the ultraviolet rest-frame--than any of the
starburst spectrophotometric templates.  In each of these cases, $\alpha <
0.5$.)  The spectral energy distributions are selected according to a
$\chi^2$ analysis similar to the analysis used for our previous photometric
redshift measurements (Lanzetta \etal\ 1996; FLY99).

  Next, we modify the fiducial spectral energy distribution of each galaxy by
adding the effect of intrinsic Lyman-limit absorption, which is characterized
by an assumed Lyman-limit optical depth $\tau_{\rm g}$ that is taken to be
universal across all galaxies. For a Lyman-limit optical depth $\tau_{\rm g}$, 
the output spectrum $f(\lambda,\tau_{\rm g})$ in terms of the input spectrum 
$f(\lambda,0)$ at rest-frame wavelength $\lambda$ is
\begin{equation}
f(\lambda,\tau_{\rm g}) = f(\lambda,0) \, \exp [-\tau_{\rm g} (\lambda/\lambda_{\rm
LL})^3]
\end{equation}
at rest-frame wavelengths $\lambda < \lambda_{\rm LL}$, where $\lambda_{\rm 
LL}$ is the rest-frame wavelength of the Lyman limit.

Next, we modify the fiducial spectral energy distribution of each galaxy by
adding the effect of random realizations of intervening Lyman series and Lyman
limit absorption.  Here we characterize the distribution of Lyman $\alpha$
forest absorption systems by a distribution function $F(N, b, z)$, which is
defined in such a way that $F(N, b, z) \; dN \; db \; dz$ is the expected
number of absorption systems in the neutral hydrogen column density interval
$dN$ around $N$, Doppler parameter interval $db$ around $b$, and redshift
interval $dz$ around $z$.  We adopt as the functional form of the distribution
function
\begin{equation}
F(N, b, z) \propto (1 + z)^\gamma N^{-\beta} G^*(\hat{b}, \sigma_b, b_{\rm
min}),
\end{equation}
where $G^*(\hat{b}, \sigma_b, b_{\rm min})$ is a ``truncated'' Gaussian
distribution of mean $\hat{b}$, dispersion $\sigma_b$, and minimum truncation
value $b_{\rm min}$.  We adopt $\gamma = 2.5$, $\beta = 1.6$, $\hat{b} = 23$
km s$^{-1}$, $\sigma_b = 8$ km s$^{-1}$, and $b_{\rm min} = 15$ km s$^{-1}$.

We have extensively checked that simulated spectra generated according to
these parameters reproduce to within observed scatter the average Lyman
$\alpha$ absorption $D_A$ and average Lyman series absorption $D_B$. In
Figure 1 we show the average values of $D_A$ and $D_B$, compared to values
found in the literature. The agreement is perfect within the scatter present
in the observational data.

The incidence of Lyman limit systems (those with hydrogen opacity $\tau > 1$)
is an important factor in our simulation, as they are amongst the main
drivers of the absorption bluewards of the galactic Lyman edge. We plot in
Figure 2 the density of Lyman limit systems generated by our Lyman $\alpha$
forest model, and compare it with the numbers measured by different authors
(as described in the Figure). As can be seen, our model produces slightly
less Lyman limit systems at the lowest redshifts than have been observed.
Whereas this effect could produce a bias favouring lower escape fraction
values, we have checked that this effect is not important (see \SS 4.2 and
4.3 for details). This slight deffect is the result of our decission to adopt
a single parameterization to generate all of the Lyman $\alpha$ forest--a
change in the normalisation to improve the aspect of Figure 2 would conflict
with the data shown in Figure 1.  Globally, we consider that our model
reproduces within reasonable limits the properties of the intergalactic
medium.

\begin{figure}
\null \hfil
\psfig{figure=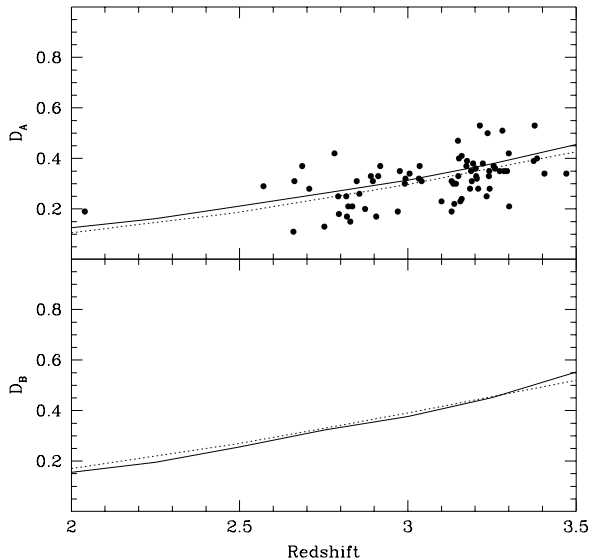,width=80mm} \hfil \null
\caption{Values of the Lyman decrements $D_A$ and $D_B$ obtained by averaging
  1000 spectra generated by our model at each redshift (continuous line). The
  curves are compared with the values by Webb (unpublished, dotted line)
  which our group uses for photometric redshift estimation, and with a
  compilation of $D_A$ values obtained from the literature (Oke \& Korycansky
  1982; Bechtold \etal\ 1984; Steidel \& Sargent 1987; O'Brian, Wilson \&
  Gondhalekar 1988; Schneider, Schmidt, \& Gunn 1989; Giallongo \& Cristiani
  1990; Schneider, Schmidt, \& Gunn 1991).}
\end{figure}

\begin{figure}
\null \hfill 
\psfig{figure=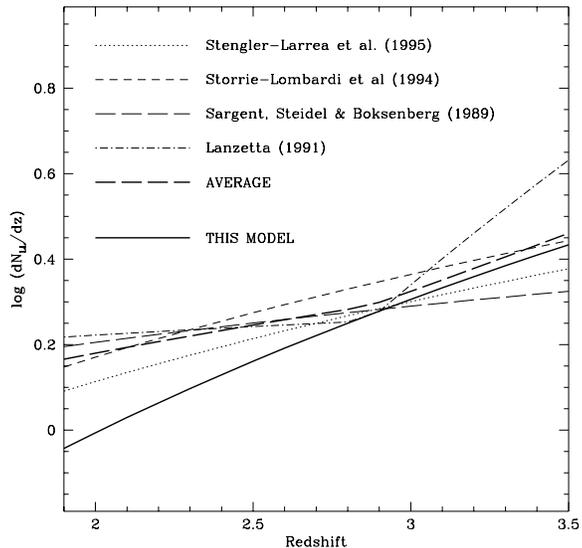,width=80mm} \hfill \null
\caption{Number density of Lyman limit systems as observed by several authors
 (see labels) and produced by our method. The difference between our model and
 the average of the different observations is always less than 0.2dex in the
 redshift range of interest, and almost zero at $z\ge2.8$.}
\end{figure}

  Next, we measure the simulated spectrum of each galaxy---constructed from
the fiducial spectral energy distribution of each galaxy modified by the
effects of intrinsic Lyman-limit absorption with an assumed Lyman-limit
optical depth $\tau_{\rm g}$ and of random realizations of intervening Lyman
series and Lyman limit absorption---at short-wavelengths, i.e.\ over the F300W
bandpass which is sensitive at wavelengths below the rest-frame Lyman limit,
and we compare observed energy fluxes $f_{\rm obs}^{(i)}$ and uncertainties
$\sigma_{\rm obs}^{(i)}$ and simulated energy fluxes $f_{\rm sim}^{(i,j)}$ of
the $i$th galaxy in the $j$th simulation.  To randomize the effects of
intervening Lyman series and Lyman limit absorption, we repeat this procedure
a large number of times, and we form the likelihood as a function of
Lyman-limit optical depth $\tau_{\rm g}$ as
\begin{equation}
{\cal L}(\tau_{\rm g}) = \prod_{i=1}^{N_{\rm gal}}
\left(
\frac{1}{N_{\rm sim}}
\sum_{j=1}^{N_{\rm sim}}
\frac{\exp \left[
-\frac{(f_{\rm obs}^{(i)}-f_{\rm sim}^{(i,j)}(\tau_{\rm g}))^2}
{2 {\sigma_{\rm obs}^{(i)}}^2}
\right]}
{\sqrt {2 \pi {\sigma_{\rm obs}^{(i)}}^2}}
\right),
\end{equation}
where the product extends over the $N_{\rm gal}$ galaxies of the sample and
where the sum extends over the $N_{\rm sim}$ simulations of the analysis.  We
find that $N_{\rm sim} = 100$ is enough to represent the variation of
intervening Lyman series and Lyman limit absorption.

\section{Results and Discussion}

  Figure 3 shows results of the method described in \S\ 3 applied to five
examples of the galaxies.  The examples are chosen to span a range in spectral
type, redshift, and luminosity.  For clarity, only 10 of the 100 realizations 
are presented for each galaxy.  It is obvious from inspection of Figure 3 that 
in many cases (especially at the lowest redshifts) the fluxes observed through
the F300W filter are less than the fluxes expected, if the galaxies were 
completely transparent to ionising photons.  This trend is also present in
the other galaxies of the sample.  Our quantitative maximum-likelihood analysis
of this trend across the entire sample of galaxies establishes a lower limit to
the optical depth at the Lyman limit or an upper limit to the escape fraction
for these galaxies.  In this section, we discuss these limits.

\begin{figure*}
\null \hfill \psfig{figure=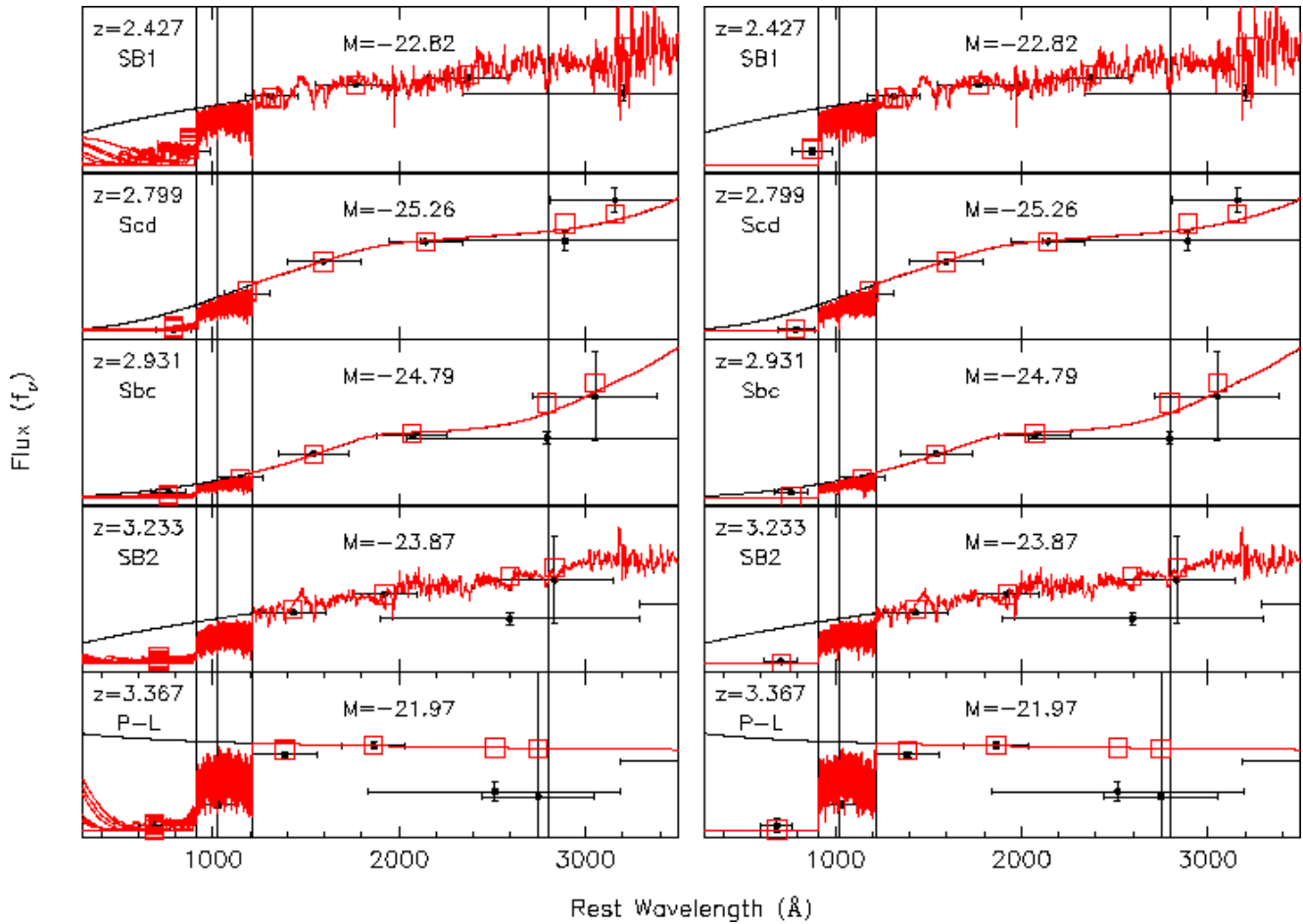,width=175mm,angle=-90} \hfill \null
\caption{Photometry and model spectra (10 simulations each time) of a sample 
of galaxies with no intrinsic Lyman limit absorption ($\tau_{\rm g}=0$, left)
and strong intrinsic absorption ($\tau_{\rm g}=10$, right). The vertical
lines mark the positions of the Lyman limit, Lyman $\beta$ and Lyman $\alpha$
lines, and $\lambda=2800$\AA\ in the rest frame. Circles with error bars show
the photometric measurements. The squares show the photometry obtained from
the fitted spectrum+HI absorption. The galaxies have been chosen to show the
full range in redshift, type, and luminosity.}
\end{figure*}

\subsection{Galaxy Escape Fraction}

  Figure 4 and Table 2 show results of the method described in \S\ 3 applied to
the data described in \S\ 2.  The first panel of Figure 4 shows the relative
likelihood as a function of $\log(\tau_{\rm g})$ determined from the entire
sample of galaxies.  The best-fit value of the galaxy Lyman-limit optical depth
determined from the entire sample of galaxies is
\begin{equation}
\tau_g = 5.2 ^{+1.4}_{-0.8},
\end{equation}
and the corresponding best-fit value of the galaxy escape fraction determined
from the entire sample of galaxies is
\begin{equation}
f_{\rm esc} = 0.008 \pm 0.006.
\end{equation}
These results apply at the average redshift of the entire sample of galaxies,
which is
\begin{equation}
\langle z \rangle = 2.77
\end{equation}
The $3\sigma$ upper limit to the galaxy escape fraction determined from the
entire sample of galaxies is
\begin{equation}
f_{\rm esc} < 0.039.
\end{equation}
Apparently, our analysis rules out the possibility of a large galaxy escape
fraction at high significance but cannot rule out (at the $3\sigma$ level) the
possibility of a galaxy escape fraction identically equal to zero.  Based 
on these results, we conclude that {\em galaxies of redshift $z \approx 3$ are 
highly opaque to ionising photons.}

\subsection{Possible selection Effects}

   Results of our analysis are apparently at odds with results of Steidel et
al.\ (2001) that galaxies of $\langle z\rangle \approx 3.4$ are not highly
opaque to ionizing photons.  These authors observed in the average spectrum
of 29 galaxies a flux ratio $F_{9/15}=0.056$.  After correcting for the
effects of the Lyman $\alpha$ forest absorption using a composite QSO
spectrum at the same redshift, they found an intrinsic flux ratio of
$F_{9/15}=0.22 \pm 0.05$--where the effects of intrinsic galactic absorption
and attenuation by dust or gas have not been eliminated.  The authors further
incorporated to this measurement a plausible estimate of the flux
discontinuity at the Lyman limit due to absorption in stellar atmospheres to
derive an escape fraction $f_{\rm esc} \ga 0.10$.\footnote{Steidel and
collaborators used a definition of $f_{\rm esc}$ which is different from the
usual one--they normalised the escape fraction at 900 \AA\ to the escape
fraction at 1500 \AA.  They reported $f_{\rm esc} \ga 0.50$, which is
converted here to the usual definition assuming the escape fraction at 1500
\AA\ is $\approx 0.15 \relbar 0.20$ as quoted by the authors.}

  Steidel and collaborators suggested that the high escape fraction found in 
their work might result from some selection bias of their galaxy sample.  The 
galaxies incorporated into their analysis include only 29 of 875 galaxies for 
which they obtained spectroscopic observations with the Keck telescope, 
selected on the basis of quality of the observations at blue wavelengths.  This
selection condition quite naturally selects galaxies that are either very 
blue, very luminous, or both.  QSOs are known to be transparent to ionizing 
photons at ultraviolet wavelengths (Zheng \etal\ 1997), so it is not 
unreasonable to speculate that very blue or very luminous (or both) galaxies 
might exhibit larger escape fractions than avarage galaxies.  To examine this 
possibilty, we repeated the analysis described in \S\ 3 for various subsamples 
of the galaxy sample.

{\em Color:}  First, we repeated the analysis for two subsamples of the galaxy
sample divided according to color.  Results are shown in the second panel of
Figure 4 and the second group of entries in Table 2.  Here ``bluer'' galaxies
are galaxies best described by SB1 or power-law spectrophotometric templates
and ``redder'' galaxies are galaxies best described by Scd, Irr, and SB2
spectrophotometric templates.  Results obtained from both subsamples are
consistent with each other and with results obtained from the entire sample of
galaxies, although there is a (statistically insignificant) trend for the
bluer galaxies to indicate a {\em smaller} galaxy escape fraction than is
indicated by the redder galaxies.  This result runs contrary to the idea
described above.

{\em Luminosity:}  Next, we repeated the analysis for two subsamples of the
galaxy sample divided according to luminosity.  Results are shown in the third 
panel of Figure 4 and the third group of entries in Table 2.  Here
``low-luminosity'' galaxies are galaxies of absolute magnitude $AB > -23.75$
and ``high-luminosity'' galaxies are galaxies of absolute magnitude $AB <
-23.75$.  Results obtained from both subsamples are consistent with each other
and with results obtained from the entire sample of galaxies, although there is
a (statistically insignificant) trend for the lower-luminosity galaxies to
indicate a smaller galaxy escape fraction than is indicated by the
higher-luminosity galaxies.

{\em Redshift:} Finally, we repeated the analysis for two subsamples of the
galaxy sample divided according to redshift.  Results are shown in the third
panel of Figure 4 and the third group of entries in Table 2.  Here
``low-redshift'' galaxies are galaxies of redshift $1.95 < z < 2.85$ and
``high-redshift'' galaxies are galaxies of redshift $2.85 < z < 3.50$.
Results obtained from both subsamples are consistent with each other and with
results obtained from the entire sample of galaxies, although there is a
(statistically insignificant) trend for the lower-redshift galaxies to
indicate a smaller galaxy escape fraction than is indicated by the
higher-redshift galaxies.  To some extent, this statistically insignificant
difference could be due to the already mentioned relative lack of Lyman limit
systems in our model at low redshift. Given that the results from the
high-redshift sample (where our model reproduces exactly the observed number
of Lyman limit systems) are not significatively different from the results
obtained using the low redshift sample, we infer that the difference in the
model does not induce a large change.

  It should be remarked that {\it none of the subsamples analysed is different
from the values given by the complete sample at the 3$\sigma$ level.}  Neither
are any of the subsamples incompatible with their complementary subsamples at
the same level.  With this in mind, it is nevertheless interesting that the
apparent effects of luminosity and ultraviolet blueness on $f_{\rm esc}$ seem
to pull in opposite directions, suggesting that the higher escape fraction
reported by Steidel and collaborators is unlikely due to selection effects.

\begin{table*}
  \centering 
   \caption{Measurements of $f_{\rm esc}$ obtained via the maximum
   likelihood method described in the text. The different subsamples used in
   the analysis are listed, together with the results and {\it statistical}
   confidence intervals associated to each one.}
   \begin{tabular}{l r c c c c}
\hline
Sample & N\footnote{Number of objects included in the sample} & $f_{\rm esc}$ &
$1\sigma$ Interval & $2\sigma$ Interval & $3\sigma$ Interval \\ \hline 
All     & 27 & 0.008 & (0.001--0.013) & (0.000--0.023) & (0.000--0.039) \\
\\
Redder  & 15 & 0.008 & (0.004--0.020) & (0.000--0.036) & (0.000--0.054) \\
Bluer   & 12 & 0.000 & (0.000--0.003) & (0.000--0.018) & (0.000--0.047) \\
\\
High Luminosity&13& 0.026 & (0.017--0.050) & (0.008--0.072) & (0.002--0.093) \\
Low luminosity &14& 0.000 & (0.000--0.001) & (0.000--0.004) & (0.000--0.013)\\
\\
Low redshift & 13 & 0.004 & (0.000--0.008) & (0.000--0.018) & (0.000--0.030) \\
High redshift& 14 & 0.169 & (0.056--0.389) & (0.008--0.857) & (0.000--1.000) \\
\hline
   \end{tabular}
\end{table*}

\begin{figure}
\null \hfill 
\psfig{figure=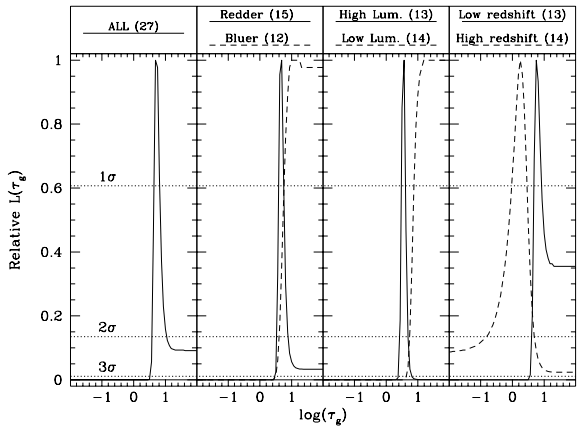,width=80mm,angle=-90} \hfill \null
\caption{Relative likelihoods of the value of $\tau_{\rm g}$ for the complete
sample (left panel). The other panels show the results for different
subsamples, broken according to (left to right) type, redshift, and
luminosity. In all panels the horizontal lines correspond to statistical 1,
2, and 3 $\sigma$ confidence intervals.}  
\end{figure}

\subsection{Systematic uncertainties}

All the confidence limits reported above refer only to the statistical
properties of the sample of galaxies we are analysing. It is very likely that
the method we are using to analyse the data may introduce errors of a
systematic nature, that could potentially be of the same order of the
statistical ones. We have performed some checks in order to estimate which
could be the contribution from several sources of uncertainty.

\subsubsection{Uncertainties induced by the Lyman $\alpha$ forest model}

As was shown in \S 3 above, our Lyman $\alpha$ forest model has been checked
against the observed properties, and agrees with them within observational
limits. In order to estimate the uncertainty in $f_{\rm esc}$ that could
originate from the uncertainties in the model, we have performed new
measurements of $f_{\rm esc}$ using forest models that differ from our
``standard'' one by applying a 10\% increase/decrease in the normalisation of
the line density. We must remark that this 10\% change does largely
overestimate the possible uncertainties in our model--the models with 10\%
more or less lines represent bad fits to the observed forest properties.

As expected, an increase in the forest density implies an increase in the
measured value of the escape fraction: more of the UV photons can escape each
galaxy while keeping the observed photometric properties, as their mean free
path is reduced by the increased density of absorbers. {\it Mutatis
mutandis}, when the forest density is reduced, the value of $f_{\rm esc}$
moves down. The change is, however, small: at the $3\sigma$ confidence level,
the original limit ($f_{\rm esc}<0.039$) becomes $f_{\rm esc}<0.030$ ($f_{\rm
esc}<0.046$) when the line density is decreased (increased) by 10\%.

Another potential problem in our forest model could be the lack of clustering
in our simulations, as opposed to what is shown by the observations. Our
measurement is not affected by the small-scale clustering that has been
observed in the Lyman $\alpha$ forest lines (Fern\'andez-Soto \etal\ 1996)
because we are integrating over much larger scales. Of potential importance
is the observed clustering of high-column density absorbers (Sargent,
Boksenberg \& Steidel 1988). It is difficult to quantify its effect, but we
expect it should be dilluted by the size of our sample, and estimate that it
produces a smaller variance than the one induced by the $\pm$10\% change in
the number of lines presented above.

\subsubsection{Uncertainties induced by the choice of the spectral templates}

In the case of some of the galaxies, the photometric data available could
allow for the assignation of two different fiducial spectral templates with
similar goodness-of-fit values. The choice of one of them over the other
(which we perform algorithmically via the minimum chi-squared criterium)
could produce an extra uncertainty in the measurement of $f_{\rm esc}$. The
same effect can be produced if our selection of templates is not dense enough
to cover the possible UV shapes, or if the photometric uncertainties are too
big.

We have evaluated the uncertainty induced by this effect by performing the
following exercise: we have forced all spectral templates to be one step
bluer (always along the sequence formed by Ell, Sbc, Scd, Irr, SB2, SB1) than
the one which is the best-fit choice\footnote{Exception made of the ones that
are already fitted by a SB1 template or a power-law}, and repeated the
analysis. This does in fact result in very bad fits for many of the galaxies,
and must thus be considered {\it a very large overestimate} of the possible
uncertainties. Under this scenario, with the galaxies emitting far more
ionising photons than what is indicated by their actual UV continua, the
escape fraction diminishes to very low values: $f_{\rm esc}<0.016$ at the
$3\sigma$ level.

We have also performed the opposite exercise, forcing all galaxies to be one
step redder than their best-fit template\footnote{Exception made of those
galaxies fitted by a power-law}. Under this assumption the galaxies emit far
less ionising photons than what their UV continua actually suggests, and
this results in an increase in the acceptable escape fraction, which reaches
$f_{\rm esc}=0.065$ , with a $3\sigma$ upper limit $f_{\rm esc}<0.156$.

As we explained before, these limits must be taken as {\it very large
overestimates} of the possible systematic effects induced by our
spectroscopic templates, as we do not realistically expect to mistake all
galaxies in the same direction. However, even under this extreme assumption,
and incorporating also the effects described in \S 4.3.1, the upper limit to
the measured escape fraction is still far below the values measured by
Steidel \etal\ (2002).

\section{Conclusions}

We have presented a new method to measure the escape fraction of ionising
photons from galaxies at $z \approx 3$. Our method provides a cleaner
measurement compared to the classical spectroscopic technique because the
problem of the sky subtraction is reduced almost to insignificance and we can
use a wider bandpass.

The value we obtain ($f_{\rm esc}=0.008$, with $f_{\rm esc}<0.039$ at a
$3\sigma$ confidence level) is in agreement with most previous measurements
of the escape fraction of ionising photons at high and low redshift, being more
stringent than most previous upper limits. It is largely in contradiction with
the value presented by Steidel \etal\ (2001) for high-redshift
galaxies\footnote{Steidel \etal\ (2001) remark that their results ``should be
treated as preliminary until high quality observations of individual galaxies
exhibiting clear evidence for Lyman continuum photon leakage become
available''}. We have studied possible selection effects that could cause
this apparent difference and conclude that, {\it within the limits set by the
size of our sample}, they cannot explain the large discordance between our
results. Neither can the difference be explained by possible systematic
effects induced by our method, as we also have shown. A (very conservative)
estimate of the confidence limits when all systematic effects are included
leads to a $3\sigma$ upper limit to the escape fraction that cannot be higher
than $f_{\rm esc} \lllla 0.15$.

If the escape fraction of ionising photons reported here represents the general
situation at high redshift, then normal galaxies cannot be responsible for any
significant fraction of the high-redshift ionising background.  On the other
hand, if the proximity effect measurements of the background flux are
confirmed, then the problem persists to find the objects responsible for the
ionisation state of the high-redshift universe. Deep, narrow-band imaging of
local and high-redshift galaxies at wavelengths slightly above and below their
intrinsic Lyman limits could settle this argument. Such imaging campaigns
would be less expensive, in terms of observing time, than the spectroscopic
observations done to date.

\section{Acknowledgments}

We thank our referee, Emanuele Giallongo, for his useful comments that have
improved the clarity of our paper. AFS gratefully acknowledges support by a
Marie Curie Fellowship of the European Community programme {\it ``Improving
the human research potential and the socio-economic knowledge base''} under
contracts number MCFI-1999-00494 and MCFI-2002-00472. HWC acknowledges the
hospitality of the Osservatorio di Brera-Merate where this work was
completed.

\label{lastpage}

\end{document}